\documentclass[12pt]{article}
\begin{document}
\begin{center}
\LARGE
\textbf{Remarks on Mohrhoff's Interpretation
of Quantum Mechanics}\\[1cm]
\large
\textbf{Louis Marchildon}\\[0.5cm]
\normalsize
D\'{e}partement de physique,
Universit\'{e} du Qu\'{e}bec,\\
Trois-Rivi\`{e}res, Qc.\ Canada G9A 5H7\\
email: marchild$\hspace{0.3em}a\hspace{-0.8em}
\bigcirc$uqtr.ca\\
\end{center}
\medskip
\begin{abstract}
In a recently proposed interpretation
of quantum mechanics, U. Mohrhoff advocates original
and thought-provoking views on space and time, the
definition of macroscopic objects, and the meaning of
probability statements.  The interpretation
also addresses a number of questions about factual
events and the nature of reality.  The purpose of this
note is to examine several issues raised by Mohrhoff's
interpretation, and to assess whether it helps providing
solutions to the long-standing problems of quantum
mechanics.
\end{abstract}
\section{Introduction}
More than 75 years after quantum mechanics was
proposed and the wave function given a probabilistic
interpretation, the issue of interpreting the
theory is anything but settled.  The past 25 years
have witnessed, among other avenues, the revival of Bohmian
mechanics~\cite{bohm1,bohm2,holland}, the development
of decoherence theory~\cite{zurek}, the consistent
histories approach~\cite{griffiths,omnes}, spontaneous
localization theories~\cite{ghirardi}, the evolution
of the modal interpretation~\cite{fraassen,vermaas},
and the idea of a veiled reality~\cite{despagnat}.

The main question all these approaches try to
answer is the following: How can one reconcile the
apparently indeterminate nature of quantum observables
with the apparently determinate nature of classical
observables, if the latter are reduced to the former?
The question shows up most spectacularly in the 
so-called quantum measurement problem, recognized
early after the advent of quantum 
mechanics~\cite{neumann}.  It is also deeply
connected with the meaning of probability statements.

A new and original way of looking at these problems
has, in the past few years, been proposed by
U.~Mohrhoff.\footnote{Most relevant to the present
discussion are Refs.~\cite{mohrhoff1,mohrhoff2,
mohrhoff3,mohrhoff4,mohrhoff5,mohrhoff6}.  Others can
be found in the e-print archives
\texttt{http://arXiv.org}.}  Like the Copenhagen
interpretation, with which it shares a number of
characteristics, it aims at ``an interpretation of quantum 
mechanics that takes standard quantum mechanics 
to be fundamental and complete.''~\cite[p.~6]{mohrhoff2}
But it goes beyond the Copenhagen interpretation
in making rather striking proposals on several most
fundamental issues, among them
the structure of space-time, the nature of physical 
reality, and the meaning of objective probability.

This paper is a critical examination of a number of
assertions made in Mohrhoff's interpretation, together
with an analysis of their consequences.  The various
questions addressed here do not cover all aspects of the
interpretation, which partakes of a wide-ranging system
that reaches well into metaphysical ontology.  The
questions I investigate, however, carry much of the 
interpretation's specific answers to
the problem of making sense of quantum mechanics.

Mohrhoff's interpretation of quantum mechanics is
briefly summarized in the following section.  Next
I examine a number of issues it raises in connection
with the Aharonov-Bergmann-Lebowitz (ABL) rule, the
quantum state, and factual reality.  I will try to
assess whether the interpretation provides
acceptable solutions to the problems of quantum
mechanics, and the extent to which the proposed
solutions are compulsory.
%
\section{Outline of Mohrhoff's interpretation}
It is impossible in just a few pages to really
do justice to Mohrhoff's rich and complex
interpretation.  This section accordingly brings
up those concepts and issues that subsequently will
be the subject of analysis.

Central to Mohrhoff's interpretation is the belief,
which he also finds in Mermin~\cite{mermin},
``that all the mysteries of quantum
mechanics can be reduced to the single puzzle
posed by the existence of objective 
probabilities.''~\cite[p.~728]{mohrhoff1}
As a matter of fact, the idea that understanding
quantum mechanics requires understanding probability
goes back at least to Popper~\cite{popper}, who
for that purpose introduced his ``propensity''
interpretation of probability.  Mohrhoff, however,
views the problem differently.  For him, objective
probabilities must be assigned on the basis of all
relevant facts, and this is done by the use of the
Aharonov-Bergmann-Lebowitz rule~\cite{abl}.

The ABL rule can be formulated as follows.  Let
$A$, $C$, and $B$ be three observables, pertaining
to a system $S$, which for our purposes can be 
taken as nondegenerate.  At time $t_1$, $S$ is prepared
in an eigenstate $|a\rangle$ of $A$.  At time $t > t_1$,
the observable $C$ is measured and one of the results
$c_j$ is obtained.  Finally, at time $t_2 > t$, $B$ is
measured and the result $b$ is found.  Then the probability
that $c_i$ is obtained at $t$, conditional upon the
preparation of $|a\rangle$ at $t_1$ and the result $b$
at $t_2$, is given by\footnote{With kets in the 
Schr\"{o}dinger picture, the Hamiltonian is assumed to
vanish between $t_1$ and $t_2$.  Equation~(\ref{abl})
also holds for a nonvanishing Hamiltonian, provided the
Heisenberg picture is used instead.}
\begin{equation}
P_{ABL} (c_i) 
= \frac{|\langle b|c_i \rangle \langle c_i|a \rangle |^2}
{\sum_j |\langle b|c_j \rangle \langle c_j|a \rangle |^2} .
\label{abl}
\end{equation}
Ket $|b\rangle$ is the eigenvector of $B$ associated
with $b$ and the sum on $j$ runs over all eigenvalues
of $C$.  Here and elsewhere, all kets are assumed
normalized.  

How can the ABL rule be used to compute objective
probabilities?  According to Mohrhoff, probabilities
of different outcomes are objective only if they are
based on all relevant facts, past, present, and future.
Thus if $C$ is measured, and the result $c_i$ is found,
then the objective probability of $c_i$ is trivially one and
the probability of $c_j$ for $j \neq i$ vanishes.
Nontrivial objective probabilities can therefore only
be assigned to measurements that are not performed.
That is, they are assigned counterfactually.  The
ABL rule provides objective probabilities if it is
interpreted as giving the probability of $c_i$ at $t$,
on the conditions that (i) $S$ is prepared in $|a\rangle$ at
$t_1$, (ii) no measurement is made between $t_1$ and 
$t_2$, and (iii) $S$ is found in $|b\rangle$ at $t_2$.

The Born probability of $c_i$ can also be
defined as $|\langle c_i|a \rangle|^2$.  But in the present
context, it is subjective, inasmuch as it does not 
take into account the (later) result~$b$.  The Born
probability is objective only if no measurement is
performed on $S$ after its preparation at $t_1$.

In general, the ABL rule (as well as the Born rule if
no measurement is performed after $t_1$)
objectively assigns nontrivial
probabilities to various possible results $c_j$.  This,
so the argument goes, has far-reaching consequences.
``The objectivity of quantum-mechanical probabilities
\mbox{[\ldots]} entails that the contingent properties
of material objects are extrinsic rather than
intrinsic.''~\cite[p.~728]{mohrhoff1}  Extrinsic properties
``supervene on what happens or is the case in the rest
of the world.''~\cite[p.~869]{mohrhoff4} In between
preparation and measurement or, more generally,
outside the context of a property-indicating fact, no
quantum observable has a value.  This 
``nonvaluedness'' holds also in the case where the
objective probability of a given result is 
one.\footnote{``Nonvaluedness'' is introduced here
with a meaning akin to Mohrhoff's ``fuzziness.''  Both
terms pertain to a quantum observable in between 
measurements.  Fuzziness, however, specifically applies to
the case where the objective probability of at least
two results differs from zero.}  That is, one cannot assign
an ontological ``element of reality'' to
a quantum observable even if its objective (ABL or Born)
probability is one.\footnote{Elements of reality were
introduced by Einstein, Podolsky, and Rosen in their
seminal 1935 paper~\cite{einstein}.  Following Mohrhoff,
I use the characterization given by
M.~Redhead~\cite[p.~72]{redhead}: ``If we can predict
with certainty, or at any rate with probability one,
the result of measuring a physical quantity at time~$t$,
then at the time~$t$ there exists an element of reality
corresponding to the physical quantity and having a value
equal to the predicted measurement result.''}

Facts, it turns out, are fundamental in 
Mohrhoff's interpretation.  A typical fact (or matter of
fact, actual event, or state of affairs) is the click
of a Geiger counter.  ``Quantum mechanics always presupposes,
and therefore never allows us to infer, the existence
of a fact that indicates the alternative
taken.''~\cite[p.~733, original
emphasized]{mohrhoff1} Facts are uncaused and
irreducible.  Quantum mechanics can, for instance,
give the subjective Born probability that $c_i$ is
found at $t$ if $S$ is prepared at $t_1$ \emph{and if}
a $C$-indicating fact occurs at~$t$.  But it cannot
predict, either categorically or probabilistically, that
such a fact will indeed occur.

The genuine nonvaluedness of quantum observables applies in
particular to space and time.  Hence it should come as
no surprise that ``[t]he problem
of understanding quantum mechanics is in large measure
the problem of finding appropriate ways of thinking about
the spatial and temporal aspects of the physical
world.''~\cite[p.~2]{mohrhoff3}  Space, for instance, has
physical reality only if evinced by a 
position-indicating fact.  In the two-slit experiment, the
conceptual distinction we make between two regions of space
does not exist for the electron unless a fact (e.g.\ a
measurement successfully performed) indicates it.
Such a fact will not in general indicate
the position with arbitrary precision. The position
of a quantum object therefore only exists when warranted
by a fact, and then only within the limit where it is indeed
warranted.  ``[T]he standard, substantival, 
set-theoretic conception of space
is as inconsistent with quantum mechanics as absolute
simultaneity is with special relativity.''~\cite[p.~3]{mohrhoff3}
Space as an infinitely-differentiated 
objectively-existing manifold where events occur is simply
a fiction.  The same applies to time.  Outside the
context of a moment-indicating fact, time does not
exist.

The nonexistence of time as an objective 
one-dimensional continuum, together with the
nonvaluedness of all quantum observables outside the
context of facts, leads Mohrhoff to conclude that
there is no such thing as an evolving quantum state.
The ``state vector'' $|\psi (t) \rangle$ (or the density
operator $\rho (t)$) is strictly
a tool used to compute either the subjective 
probability of various results at $t$ if a fact
occurs at $t$, or the objective probability of
results at $t$ if no fact occurs at $t$.  It does
not represent the time evolution of a state because
(i) a state and a probability measure are logically
different categories; (ii) there is no time $t$ 
if there is no fact at $t$;
and (iii) no observable (even one which has
$|\psi (t) \rangle$ as eigenvector) has a value unless
the value is indicated by a fact.  The last two reasons
are related, for ``[t]he insufficiency of Redhead's
`sufficiency condition' [for the existence of an
element of reality] hinges on the nonexistence of the
particular time $t$ in the absence of an actual
measurement performed at the time 
$t$.~\cite[p.~23]{mohrhoff3}
%
\section{The ABL rule}
The ABL rule was formulated in the previous section.
It turns out that the rule can be interpreted
in two very different ways.  The two interpretations
refer to different contexts, and as such are not
mutually contradictory.

Suppose that $S$ (i) is prepared in $|a\rangle$ at $t_1$,
(ii) undergoes a measurement of $C$ at $t$, and
(iii) undergoes a measurement of $B$ at $t_2$ with
the result $b$.  Following the notation of
Ref.~\cite{sharp}, we shall call this an $A \rightarrow
C \rightarrow B$ context.  In that context, the ABL
rule can be interpreted as giving the subjective
probabilities of various results of the measurement
of $C$.  That the ABL rule means at least this much
is uncontroversial.

Suppose now that $S$ (i) is prepared in $|a\rangle$ at
$t_1$, (ii) undergoes no measurement between $t_1$ and
$t_2$, and (iii) undergoes a measurement of $B$ at $t_2$ 
with the result $b$.  This we shall call an 
$A \rightarrow B$ context.  In that context, the ABL
rule can be interpreted as asserting something at
$t$ about an observable $C$ that has not been
measured.  This interpretation is controversial.  It
was not explicitly asserted by the original proponents of
the rule~\cite{abl}, although one of them later held
that they had indeed meant so~\cite{albert}.

What can be said in general about the second
interpretation?  From a purely logical point of view,
it should be clear that one cannot, from uncontroversial
factual properties of $C$ in an $A \rightarrow
C \rightarrow B$ context, infer anything factual
about $C$ in an $A \rightarrow B$ context.  For example,
if $C$ coincides with $A$, one cannot from the assertion
that $A$ is $a$ at $t$ in an $A \rightarrow
A \rightarrow B$ context infer a value for $A$ at $t$
in an $A \rightarrow B$ context.  Nevertheless,
various theories (or theory interpretations)
can make claims about ontological properties
of $C$ in an $A \rightarrow B$ context.  They can, for
instance, make assertions about the ``true value''
of $C$.  It is obvious that such claims are
absolutely untestable, since a test would involve measuring 
$C$, thereby landing in an $A \rightarrow C \rightarrow B$
context.  This, however, does not imply that these claims
have no meaning (for people other than strict empiricists,
at least) or are not interesting.

In quantum mechanics, it cannot be maintained that $C$
has a true value (equal to one of its eigenvalues)
revealed in and unaffected by a measurement.  This is a
direct consequence of the formalism and holds in any
interpretation whatsoever.  Indeed if $C$ had such a true
value unaffected by measurement, the probability of~$b$
conditional on $a$ would be given by
\begin{equation}
P(b|a) = \sum_i P(b|c_i) P(c_i|a) .
\label{prob}
\end{equation}
But in quantum mechanics $P(b|a) = |\langle b|a \rangle|^2$,
with similar formulas for $P(b|c_i)$ and $P(c_i|a)$.  These
probabilities do not, in general, satisfy~(\ref{prob}).

We should note that in an $A \rightarrow B$ context,
quantum mechanics can consistently (although not
compulsorily) be interpreted as asserting that the 
observable $A$ has a true value equal to $a$ between
$t_1$ and $t_2$.  That true value corresponds to 
Redhead's elements of reality.
Now the Born and ABL rules are symmetric
under the interchange of $a$ and $b$.  This suggests also
asserting that the observable $B$ has a true value 
equal to $b$ between $t_1$ and $t_2$.  If both assertions
are maintained, quantum mechanics must be an incomplete
theory, because if observables $A$ and $B$ do not commute,
eigenvectors of one are not in general eigenvectors of
the other.  We shall see, however, that there are good
(though again not compelling) reasons for maintaining that 
$A$ has and $B$ does not have
a true value between $t_1$ and $t_2$.

In hidden variable theories, nothing in principle 
prevents $C$, in an $A \rightarrow B$ context, from having
a true value at all times between $t_1$ and $t_2$.  Take
for instance Bohmian mechanics.  If $|a \rangle$ and
$|b \rangle$ are not position eigenstates, then $A$ and
$B$ do not commute with the position operator $X$.  Yet
the particle's position is well-defined at all times
between $t_1$ and $t_2$.  In an $A \rightarrow X
\rightarrow B$ context, however, positions would be 
different.  The measurement of $X$ at $t$ would
from then on change the particle's wave function, and
therefore its Bohmian trajectory.

Let us now examine how does Mohrhoff's interpretation
of quantum mechanics fit in the present discussion
of the ABL rule.  In his exchange with 
R.~E.~Kastner~\cite{mohrhoff4, kastner1}, Mohrhoff explains
how to quantitatively define the counterfactual
meaning of the ABL rule.  To paraphrase him, saying
that $P_{ABL} (c_i)$ is the objective probability
with which $c_i$ would be obtained, given preparation
$|a\rangle$ and measurement $b$ at $t_1$ and $t_2$,
is ``in all relevant aspects exactly the same as saying
that'' $P_{ABL} (c_i)$ is the subjective probability
with which $c_i$ is obtained given the same outcomes
at $t_1$ and $t_2$~\cite[p.~868]{mohrhoff4}.  The relevant
aspects here are the quantitative ones.  As pointed
out in section~2, however, it is clear that Mohrhoff has
much more to say about the ontological consequences of
the counterfactual meaning.  The implications are
that the observable $C$ categorically has no value
between $t_1$ and $t_2$ in an $A \rightarrow B$
context.  Recall that an observable has a value only when
the value is ascertained by a fact.  The subjective
(uncontroversial) meaning of the ABL rule, which ``in all 
relevant [quantitative]
aspects'' coincides with the objective
(counterfactual) meaning, is held to entail an objective 
nonvaluedness of $C$ in the $A \rightarrow B$
context.  It should be stressed once more that the
nonvaluedness does not logically follow from the
counterfactual meaning of the ABL rule.  It is,
nonetheless, a possible (and certainly a significant)
interpretation based upon it.

The ABL rule, written as in~(\ref{abl}), appears invariant
under time reversal.  Indeed let the signs of $t_1$, $t$,
and $t_2$ be changed and let bras be transformed into kets
under time reversal.  Then the ABL rule is transformed into
itself.  But suppose the process of measurement, instead
of occurring at a single instant, is spread over some
interval of time.  The time-reversal invariance of the
ABL rule then depends on the time-reversal properties
of the measurement process.  This has been stressed by 
L.~Vaidman~\cite{vaidman1} as far as the $C$ measurement 
goes.  Similar considerations can be made for the
initial preparation ($a$) and final measurement ($b$).
In the case where the time reversal of preselection
is postselection and where the $C$ measurement is
time-reversal invariant, then the ABL rule is also
time-reversal invariant.  But this does not have to
be so.  Take for instance von Neumann's theory of
measurement~\cite{neumann}.  In that approach,
a measurement is an interaction followed by a collapse,
a complex process that can be symbolized as
$\mbox{Int} (C) \rightarrow (C:c_i)$.
The $A \rightarrow C \rightarrow B$ process can then 
schematically be represented as
\begin{equation}
\mbox{Int} (A) \rightarrow (A:a) \rightarrow
\mbox{Int} (C) \rightarrow (C:c_i) \rightarrow
\mbox{Int} (B) \rightarrow (B:b) .
\end{equation}
This is certainly not time-reversal invariant.
For the purpose of computing probabilities of
measurement results at time $t$, state $|a\rangle$
and state $|b\rangle$ are equally useful.  Fot the
purpose of making ontological statements, however,
it is more natural in the absence of a $C$
measurement to hold that $|a\rangle$,
rather than $|b \rangle$, is the intermediate state,
since $|b \rangle$ obtains only after an
intervening interaction.

In Mohrhoff's interpretation, however, the
time-reversal invariance of the ABL rule can be held
consistently.  The reason is that in this framework,
measurements should not be analyzed, as they are
irreducible facts.  This raises questions of its own,
to which we shall come back later on.
%
\section{The quantum state}
I have already pointed out that Mohrhoff
categorically denies that the ``state vector''
$|\psi (t) \rangle$ represents an evolving
quantum state.  ``The idea that what by definition
is a tool for assigning probabilities to
\emph{possibilities} also describes an \emph{actual}
state of affairs, is simply a category
mistake.''~\cite[p.~22]{mohrhoff3}  Most people will
agree, so the argument goes, that $|\psi (t) \rangle$ 
represents a probability measure, that it yields 
the probability of various results on the condition 
that a measurement is performed.  But if no 
measurement is performed, no observable has a value.
A state and a probability measure are two entirely 
different objects.  Identifying one with the other is 
therefore logically inconsistent.

I shall argue that such a strong claim is 
unfounded.  Nevertheless, it is entirely
consistent (though not compulsory) to
deny the existence of an evolving quantum state.
This is in fact an interpretative statement.
Although it does bring substantial benefits,
it also carries problems of its own.

The greatest benefit reaped in denying the existence
of an evolving quantum state is that the
ominous quantum measurement problem is, if not
completely solved, at least considerably attenuated.
In broad outline the
measurement problem goes as follows.
An observable $A$ with eigenvectors $|a_i\rangle$
and (say) nondegenerate eigenvalues $a_i$, 
pertaining to a quantum system $S$, is subject 
to a measurement by an apparatus $M$.  The
apparatus is also to be treated quantum
mechanically.  It starts in some initial state
$|\alpha_0 \rangle$.  Suppose the quantum system
is prepared in a state $|a_j\rangle$.  A reliable 
apparatus should be built so that after suitable
interaction between $S$ and $M$ (associated with
a unitary operator $U$), the ``pointer'' of
the apparatus shows a characteristic value
$\alpha_j$, corresponding to a state
$|\alpha_j \rangle$.  Readings $\alpha_i$
and $\alpha_j$ ($i \neq j$) are unambiguous
provided that $|\alpha_i \rangle$ and
$|\alpha_j \rangle$ are macroscopically
distinct.  But now suppose that the quantum
system is prepared in a nontrivial superposition
$\sum_k c_k |a_k \rangle$ of eigenstates of
$A$.  Then the linearity of the evolution
operator implies that after interaction,
the joint quantum state of the system and
apparatus is given by
\begin{equation}
|\chi \rangle = \sum_k c_k |a_k \rangle
\otimes |\alpha_k \rangle .
\label{joint}
\end{equation}
This result seems to be flatly ruled out by
experiment, inasmuch as no apparatus is ever observed
in a superposition of macroscopically distinct
pointer states.

To derive this contradiction, it is crucial
to view the state vector~(\ref{joint}) as
indeed representing a state, that is, an
objectively existing state of affairs.  Only
then has the experimental observation
any relevance in ruling it out.  If instead
$|\chi \rangle$ represents a probability measure
for results $\alpha_k$ on the condition that
one of them shows up, the contradiction
is removed.

There still remains, however, what is known as
the ``pointer problem.''  In~(\ref{joint}), the
$|\alpha_k \rangle$ make up an orthonormal set
of eigenvectors of an observable in the apparatus's
Hilbert space, which may be denoted by~$\mathcal{A}$.
Let $\{ |\beta_l \rangle \}$ be another orthonormal set,
related to the first one by some unitary operator.
The $|\beta_l \rangle$ can be thought of as eigenvectors
of some (nonunique) operator $\mathcal{B}$.
One can write
\begin{equation}
|\chi \rangle = \sum_l c_l ' |b_l \rangle
\otimes |\beta_l \rangle .
\label{jointb}
\end{equation}
The $|b_l \rangle$ are called ``relative
states.''~\cite{everett}  They are (in general
nonorthogonal) linear combinations of the
$|a_k \rangle$.

The pointer problem is encapsulated in the following
question: What, from a fundamental point of view,
makes vectors $|\alpha_k \rangle$ more relevant 
than vectors $|\beta_l \rangle$?  Decoherence
theory~\cite{zurek} attempts to answer the question by
bringing up effects of the environment.  Barring
that, the $|\alpha_k \rangle$ are singled out
by the following mathematical criterion: in general 
they are the only ones for which the relative states
$|a_k \rangle$ are orthogonal\footnote{It is 
well-known that the biorthogonal
decomposition~(\ref{joint})
is unique if all $c_k$ have different norms.} and,
therefore, are the eigenstates of a Hermitian operator.

This criterion may be related to what Mohrhoff
has in mind when he states that Born probabilities are
conditional, among other things, on ``the observable
[$A$] that is being measured.''~\cite[p.~26]{mohrhoff3}
Whether this solves the pointer problem is, however,
far from carrying consensus.  Mohrhoff's interpretation
does not seem to bring any additional insight to
the issue.

To sum up, denying that the state vector
represents an evolving quantum state considerably
weakens the measurement problem, in a consistent way.
Nevertheless, we shall see that from a logical point
of view, this stand is by no means compelling.

It is true that a state description and a
probability measure specify distinct classes of
objects.  But these classes are not
mutually exclusive, anymore than in mathematics,
for instance, the class of topological spaces 
excludes the class of groups.  Perhaps one can argue
that it is intuitively strange to endow a probability
measure with properties of a state description.  The
converse, however, is not strange at all.  Indeed
there are good reasons to believe that if the state
(i.e.\ the actual state of affairs) is known, predictions
of some sort can be made as regards results of
measurements.  This is trivially true, for example, in
classical mechanics.  There the state is specified by
a point in phase space at some time $t$.  From this
specification, a (trivial) probability measure can
be defined on possible states at time $t'$.

Nor does the requirement of collapse, in theories
or interpretations in which it occurs,
make the notion of an
evolving state inconsistent.  Clearly, if
one is willing to contemplate deviations from
Schr\"{o}dinger evolution, numerous collapse models
are possible that may have various degrees of likeliness.
But even if one insists that the Schr\"{o}dinger equation
holds exactly, behavior akin to collapse is possible,
as Bohmian mechanics clearly, and decoherence theory
more controversially, illustrate.

I should point out that there are good
(though again not compelling) reasons to view 
the state vector as specifying an evolving state.
Let system $S$ be prepared at time $t_1$ in the state
$|\psi (t_1) \rangle$.  The preparation being
ascertained by a fact, it is agreed that at time
$t_1$, $S$ possesses the property associated with
the projector $|\psi (t_1) \rangle \langle \psi (t_1) |$.
Assume the system's Hamiltonian is known in the
time interval ($t_1, t_2$).  Then in principle,
the state vector $|\psi (t) \rangle$ is also known
at any $t$ between $t_1$ and $t_2$.  Hence one can 
predict with certainty that a measurement, at time $t$,
of the property associated with the projector
$P_{\psi(t)} = |\psi (t) \rangle \langle \psi (t) |$
will yield the value one.  In Redhead's language,
$P_{\psi(t)} = 1$ is a (time-dependent) element of 
reality.  This possibility of correctly predicting
the value of a time-dependent dynamical variable
motivates the association of the state
vector with an actual state of affairs.\footnote{``As
my aim here is to make sense of \emph{standard} quantum
mechanics, I take the nonexistence of hidden variables
for granted.''~\cite[p.~16]{mohrhoff3}  If elements of
reality are reckoned to be outside the formalism of
quantum mechanics, this statement eliminates them
tautologically.  What is really significant, I believe,
is that elements of reality and hidden variables can
be introduced while in no way changing the 
Schr\"{o}dinger equation, the association of
measurement results with eigenvalues of Hermitian
operators, or the Born rule.}

Mohrhoff rightly points out that no detector is
100\%\ efficient.  The same applies to preparation
devices.  A preparation device may not always
prepare anything, or it may prepare something
slightly different from what is intended.
Thus in the real world one may be wrong when one
assigns an element of reality.  Yet one can consistently
maintain that there is an element of reality in, say, 
99\%\ of cases.  In investigating the extent
to which facts do or do not have causes, it may
help to remember that the inefficiency of preparation
devices is irrelevant to the existence or nonexistence
of elements of reality.
%
\section{Macroscopic objects and facts}
In Mohrhoff's interpretation, the formalism
of quantum mechanics is taken to hold exactly.
Hence it should apply to everything, including
macroscopic objects.

How are macroscopic objects to be defined?
We have seen that no object (whatever its size)
has a position unless that position is indicated
by a fact.  A fact will occur in relation with a
detector that can monitor the object's position 
up to a certain precision.  Now the bigger an
object is, the slower the wave packet associated
with its center-of-mass position will spread (the
spreading goes roughly as $m^{-1}$, where $m$ is the
object's mass).  Hence for large objects,
center-of-mass positions (or indeed center-of-mass
positions of their parts) have very little time
to become appreciably fuzzy between two successive
position-indicating facts.  That very small fuzziness,
to show itself, requires a much bigger detector that 
can monitor positions with very high precision.  The
bigger detector's position still has a residual,
exceedingly small fuzziness, which however can only be
revealed by still another, this time exceedingly large,
detector.  Clearly, in the actual world, this process
cannot go on forever.  Detectors cannot be arbitrarily
large, if only to be stable against gravitational collapse.
The upshot is that there are objects so large that no
detector exists to reveal the fuzziness of their position.
These are, by definition, macroscopic objects.

It is experimentally well established that, with one type
of exception, bulk properties of
macroscopic objects follow the laws of 
classical mechanics.  The exception occurs, of course, when
the macroscopic object is used as a pointer to indicate
the unpredictable value of a microscopic object's
observable.  But barring that, classical mechanics provides
an entirely adequate description of 
things like footballs and planets.
Now these are also subject to the laws of quantum mechanics,
inasmuch as the latter apply to everything.  Can the two
laws be made consistent?  There are strong (although not yet 
entirely compelling, see~\cite{omnes}) indications that
if a macroscopic object is treated quantum mechanically,
then quantum observables can be defined whose probabilistic
behavior follows very closely the classical behavior of
associated classical observables.  Is this enough to conclude
that classical mechanics, now understood as the causal
description of the succession of facts, is reducible to
quantum mechanics?  Not so for Mohrhoff.

The reason is that quantum mechanics never predicts
the occurrence of a fact.  On the basis of facts, it predicts
the probability of various results \emph{if} an additional 
fact occurs.  But then, what is the precise nature of 
a fact?

Mohrhoff does give some answers to this question.  He 
points out that facts play in quantum mechanics a role
similar to initial conditions in classical mechanics.
``While in classical physics actuality attaches itself
to a nomologically possible world trivially through the
initial conditions, in quantum physics it `pops up'
unpredictably and inexplicably with every
property-defining fact.''~\cite[p.~729]{mohrhoff1}
As initial conditions are always assumed and never
explained, so should facts be.  ``[T]he existence
of facts---the factuality of events or states of
affairs---cannot be accounted for, anymore than we can
explain why there is anything at all, rather than
nothing.''~\cite[p.~221]{mohrhoff5}

It would seem that in some circumstances, the
occurrence of a fact can be predicted, if not through
quantum mechanics, at least empirically.  Take for
instance a Geiger counter together with a device
that can prepare a charged particle with a position
and speed as well-defined as Heisenberg's principle
allows.  At $t_1$ the particle is emitted towards the
counter with a speed set so that it arrives at
$t_2$.  Here $t_1$ and $t_2$ are defined
in terms of nearby macroscopic clocks.  With this
setup one can predict, with a high degree of 
probability, that the counter will click at $t_2$.

It turns out, however, that things are more
complicated.  The situation just described does not
evince a unique fact.  Before and after the click
one can, within the finite time differentiation
allowed by available clocks,
repeatedly and positively ascertain that there
is no click.  All these ``nonclicks'' are facts.
When the charged particle enters the Geiger counter,
the particle's and counter's wave functions become
entangled, and there is a growing probability for 
click rather than nonclick.  I agree with Mohrhoff
that standard quantum mechanics is of no help in
predicting when precisely will the click occur.

It is entirely consistent to view the occurrence
of facts as uncaused and unpredictable.  From an
epistemological point of view, however, it is risky.
Classical mechanics does not explain initial
conditions, but it explains everything else.  That
is, the totality of the world at one instant is
unexplained, but the totality of the world at all
other instants is explained.  Mohrhoff tells us that
in quantum mechanics, all the facts that constantly
betoken positions of macroscopic objects should be
taken as unexplainable.  It seems that the
unexplained here far exceeds what it is in classical
mechanics.  Inevitably, people will look for 
regularities in the occurrence of facts (beyond the
consistency of classical trajectories with
quantum-mechanical probabilities).  In a sense,
spontaneous localization theory~\cite{ghirardi}
may be viewed as doing just that.  It may or may not
be successful.  But, as pointed out 
in Ref.~\cite{stapp},
any theory that would correctly account for the
occurrence of facts would, other things being equal,
have a head start over one that does not.

Other properties of facts may be subject to 
experimental investigation and, therefore, call for
theoretical analysis.  A fact occurs every time when, 
for instance, a detector clicks or a pointer deflects.
It is crucial to observe that the fact does not
coincide with the pointer's 
position.~\cite[p.~31]{mohrhoff3}  The latter is,
ultimately, a quantum observable subject to a fuzziness
that is just too small for a suitable detector to evince
it.  But the former is not a
quantum observable (it allows assigning values to
quantum observables), and it is not subject to 
fuzziness.  Now it will presumably be agreed that 
a 1~kg detector can register a fact.  What about smaller
detectors?  Can a 1~g detector register a fact, or a
1~$\mu$g detector, or a smaller one?  Is a detector required
to have some minimum mass to be able to work as a detector?
If the answer is yes, that mass ought to be open to
investigation.

But perhaps the answer is no.  Maybe even atoms or
elementary particles can work as detectors.
Then another question comes up.  Consider a hydrogen
atom in a $2p$ state, confined in a high vacuum.
In a split second the atom is very likely to emit a
1216~\AA\ photon and fall to the ground state.
Perhaps one will maintain that when the photon has 
gone far enough, a fact has occurred.  But then the
question is, When did the fact occur?  It won't help
to say that time for the atom is not defined outside
the context of the fact, for the occurrence can be
referred to a nearby macroscopic clock.  A matter
of fact ``is something that cannot be undone or
`erased'.''~\cite[p.~18]{mohrhoff2}  If it is agreed
that the fact has already occurred at $t = 1 \, \mbox{s}$
(as shown by the macroscopic clock), it must have
occurred at some instant between $t = 0$
and $t = 1 \, \mbox{s}$.  That instant is defined
as the one starting from which it is impossible to 
undo or erase.  Some way or other, it
ought to be open to experimental investigation.
%
\section{Discussion}
The interpretation of quantum mechanics proposed
by U.~Mohrhoff bears resemblance to the
Copenhagen interpretation.  It takes quantum
mechanics to be fundamental and complete, and it
requires the validity of classical mechanics for
its formulation.  In both the state vector is a tool
for calculating probabilities.  Yet Mohrhoff's
interpretation goes beyond the Copenhagen
interpretation in several ways.  It applies not only
to measurements in the strict sense, but to all
property-indicating facts.  Moreover, it does not
share the Copenhagen interpretation's strict
instrumentalism.  Far from
remaining silent about the behavior of quantum
observables in between measurements, it explicitly
asserts their fuzziness or nonvaluedness or,
equivalently, the meaninglessness of their having
a value.  That nonvaluedness applies most
importantly to space and time, which are undefined
outside the context of measurements or facts.  ``The
seemingly intractable problem of understanding
quantum mechanics is a consequence of our dogged
insistence on obtruding onto the world, not a
spatiotemporal framework, but a spatiotemporal
framework that is more detailed than the
world.''~\cite[p.~742]{mohrhoff1}

Probability statements have a very specific meaning
in Mohrhoff's interpretation.  Objective statements
apply counterfactually to measurements that are not
performed.  Subjective statements are rational guesses
made on the basis of incomplete information, in
particular if the result of an actually performed
measurement is not known. 
The ``state vector'' is viewed strictly as a probability
measure, and emphatically not as representing an
evolving quantum state.  This goes a long way towards
solving the quantum measurement problem.  It remains,
however, an interpretational statement, not logically
implied by the formalism of quantum mechanics.  The
same holds for the rejection of elements of reality
and the assertion of an objective nonvaluedness in
between measurements or facts.

Mohrhoff presents a view of
macroscopic objects that carries much appeal.
To them quantum mechanics
applies universally and exactly.  Their positions
are quantum observables, qualitatively subject to 
fuzziness just like observables associated with 
microscopic objects.  Quantitatively, however,
the fuzziness is exceedingly small, in effect so 
small that there are no detectors large enough to
evince it.  That view of macroscopic objects
should, I believe, be investigated further.  The
notion that an object's position always shows up
through another object's fuzzy position may
entail consequences otherwise not so easily
uncovered.

Outside the explanatory scheme of quantum
mechanics are facts, which are uncaused and
unpredictable.  Quantum mechanics predicts the
probability of measurement results on the condition
that a fact occurs, but it says nothing about the
occurrence itself.  This raises questions about
the nature of facts and the extent to which they can
be experimentally investigated as well as
theoretically analyzed.  These questions, although
framed in a different setup, are related to the
ones often asked about the frontier between the
quantum and the classical.

In the semantic view of scientific
theories~\cite{fraassen}, a theory is identified with
the class of its models or interpretations.  Each
model answers the question, How can the world be the
way the theory says it is?  The models collectively
give meaning to the theory.  In the interpretation
he has put forth, Mohrhoff has shown us a 
thought-provoking and original view of the way that,
according to quantum mechanics, the world can be.
\section*{Acknowledgments}
I am grateful to Ulrich Mohrhoff for a
stimulating exchange of views.
%

%
\end{document}